\documentclass[10pt, conference]{IEEEtran}

\usepackage{float}
\usepackage[utf8]{inputenc}
\usepackage[T1]{fontenc}
\usepackage[english]{babel}
\usepackage{graphicx}
\usepackage{subfig}
\usepackage{fullpage}
\usepackage{eso-pic}
\usepackage{amsmath}
\usepackage{amssymb}
\usepackage{hyperref}
\usepackage{xspace}
\hypersetup{colorlinks=true, pdfstartview=FitV, linkcolor=red, citecolor=black, plainpages=false, urlcolor=blue}
\usepackage[all]{hypcap}
\usepackage{url}
\usepackage{amsfonts}
\usepackage{listings}
\usepackage{lipsum}
\usepackage{kantlipsum,cuted}
\usepackage{hyperref}
\usepackage{breqn}

\newcommand{\NZERO}{\ensuremath{\mathit{N0}}\xspace}

\newcommand{\CNZERO}{\ensuremath{\mathit{C\!/\!N0}}\xspace}
\DeclareMathOperator{\sinc}{sinc}
\DeclareMathOperator{\dBHz}{dBHz}
\DeclareMathOperator{\Hz}{Hz}
\DeclareMathOperator{\chips}{chips}
\DeclareMathOperator{\chip}{chip}
\DeclareMathOperator{\ms}{ms}

\begin{document}

\title{
Convolutional Neural Network for \\ Multipath Detection in GNSS Receivers
}
\author{\IEEEauthorblockN{Evgenii Munin\IEEEauthorrefmark{1},
Antoine Blais\IEEEauthorrefmark{1} and Nicolas Couellan\IEEEauthorrefmark{1}\IEEEauthorrefmark{2}}
\IEEEauthorblockA{\IEEEauthorrefmark{1}ENAC, Université de Toulouse\\
7 Avenue Édouard Belin, CS 54005, 31055 Toulouse Cedex 4, France\\
Email: \{evgenii.munin, antoine.blais, nicolas.couellan\}@enac.fr}
\IEEEauthorblockA{\IEEEauthorrefmark{2}Institut de Mathématiques de Toulouse; UMR 5219, CNRS, UPS,  Université de Toulouse\\
F-31062 Toulouse Cedex 9, France}}

\maketitle
%\ClearShipoutPicture
%\thispagestyle{plain}
%
%\clearpage

\begin{abstract}

Global Navigation Satellite System (GNSS) signals are subject to different kinds of events causing significant errors in positioning. This work explores the application of Machine Learning (ML) methods of anomaly detection applied to GNSS receiver signals. More specifically, our study focuses on multipath contamination, using samples of the correlator output signal. The GPS L1 C/A signal data is used and sourced directly from the correlator output. To extract the important features and patterns from such data, we use deep convolutional neural networks (CNN), which have proven to be efficient in image analysis in particular. To take advantage of CNN, the correlator output signal is mapped as a 2D input image and fed to the convolutional layers of a neural network. The network automatically extracts the relevant features from the input samples and proceeds with the multipath detection. We train the CNN using synthetic signals. To optimize the model architecture with respect to the GNSS correlator complexity, the evaluation of the CNN performance is done as a function of the number of correlator output points. \\

\end{abstract}


%--------------------------------------------------------------------------------------------------------------
\section{Introduction}

\par The main motivation behind the present work is to apply machine learning techniques to predict the errors caused by the multipath effect in the Global Navigation Satellite System (GNSS) using the information extracted from the correlator output of the tracking loops of a GNSS receiver.
\par
\begin{figure}
    \centering
    \includegraphics[scale=0.3]{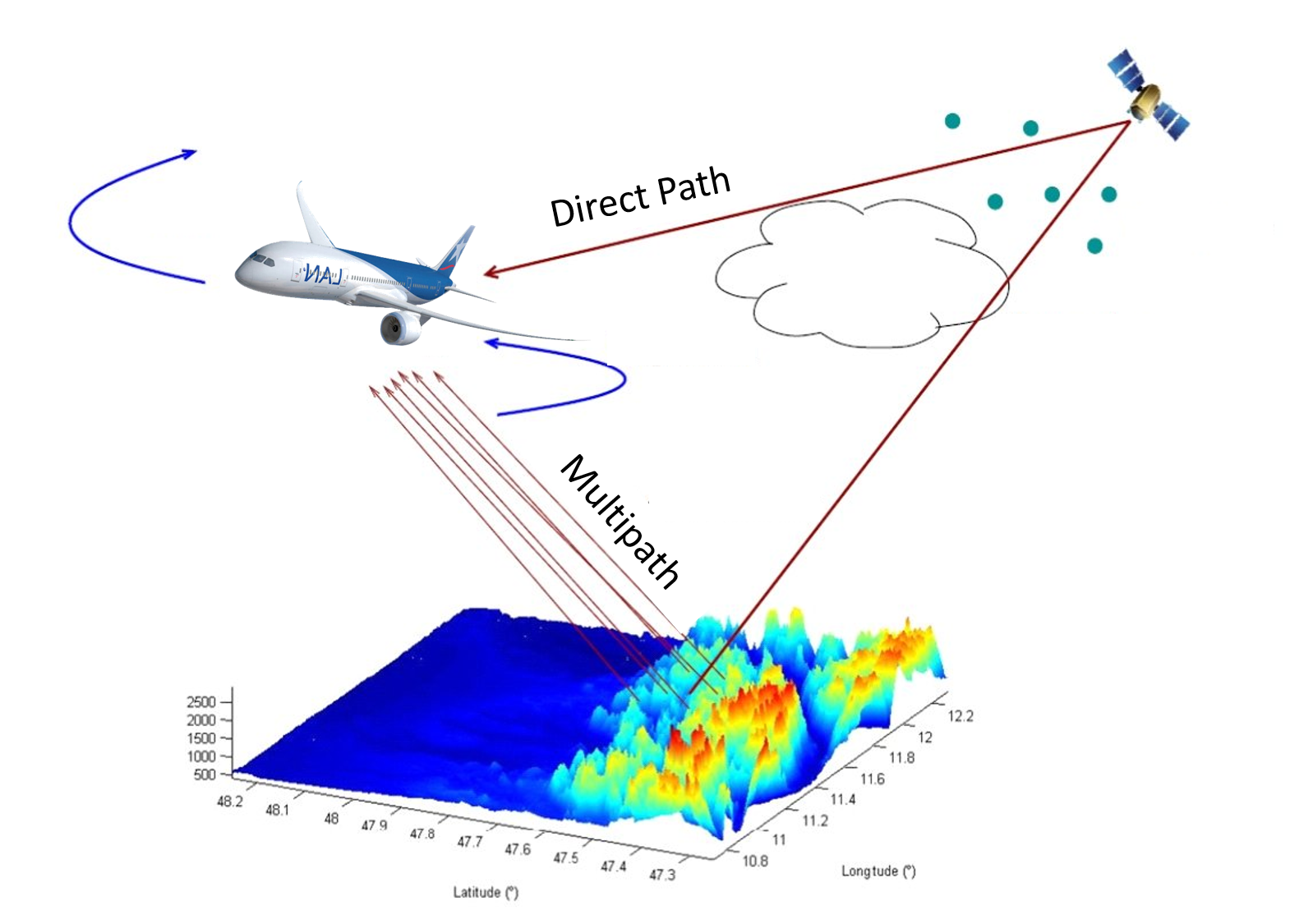}
    \caption{Multipath illustration example}
    \label{mp_example}
\end{figure}

 As mentioned in \cite{vigneau}, the multipath effect can be the source of strong deterioration of GNSS positioning performance and may be considered as the major cause of errors in urban environment \cite{urban_canyon}. It disturbs the useful signal and provides negative effect on final precision of the position delivered to user. This is the main motivation for its detection and mitigation. Multipath is defined as one or more indirect replicas of signals from satellites arriving at a receiver's antenna from a satellite as shown for example in Figure~\ref{mp_example} for the case of an aircraft.
\par To address the problem of multipath detection, we propose a solution using raw data from the correlator output represented as a tensor of images from doppler and code delay offsets. We then use a CNN to automatically extract the features and a binary multipath classifier at the last layers of the network.
\par To assess the performance of our detector we run comparison experiments with another machine learning based method using a  predefined feature construction procedure and a Support Vector Machine (SVM) classifier \cite{suzuki}.
\par The article is structured as follows. First, we review the related research done on the subject. Then we describe the GNSS correlation process and the CNN concept. Thereafter, the data generation pipeline is detailed. The experiment setup is subsequently described and finally the results are discussed.
\section{Related research}
\subsection{GNSS multipath detection and mitigation}
\par Two main approaches have been proposed in the literature, frequency-domain or time domain processing \cite{phan}. Among frequency-based approaches we can refer for example to the ones using the Fast Fourier Transform (FFT) \cite{phan-7}, or the wavelet decomposition \cite{phan-8, phan-9}. However, according to \cite{phan}, these methods can unintentionally rule out other coexisting signals of interest. In the time-domain methods we can cite the Carrier Smoothing Filter (CSF) \cite{phan-12,phan-13} and the stacking technique in the case of static receiver \cite{phan-16}. These methods also tend to remove coexisting useful signals and their performance depends on the rate of the data flow.
\par The multipath mitigation can be done at the hardware level as well as the software level. At the hardware level, the use of high quality antenna arrays can be efficient for detecting and mitigating multipath \cite{cheng-3}. They can also be used for estimating parameters of multipath. Alternatively, the multipath can be mitigated at the software level of the receiver either in acquisition or in tracking loops. As the GNSS receiver has to track the direct signal contaminated by delayed reflections, several multipath mitigation methods based on the narrow correlator Delay Lock Loop (DLL) were listed in \cite{cheng}. They include the strobe correlator \cite{cheng-6}, the early-late-slope technique \cite{cheng-7}, the double-delta correlator \cite{cheng-8} and the multipath intensive delay lock loop \cite{cheng-9}. On the other hand, a statistical approach based on the maximum likelihood principle \cite{cheng-10} or the Bayesian technique \cite{cheng-15} can be used.
\par In the case of Non-Line-of-Sight (NLOS) effect, multipath can be hardly mitigated by these approaches. To overcome these limits, \cite{cheng} proposes to use a Generalized Likelihood Ratio Test (GLRT) \cite{cheng-27} and also a Marginalized Likelihood Ratio Test (MLRT) \cite{cheng-28}, for fault detection and diagnosis.
\par To avoid the limitations of these classical signal processing methods we consider the application of Machine Learning techniques \cite{mueller2018machine, hastie01statisticallearning} to this multipath issue in GNSS. 
\subsection{Machine learning application in GNSS}
\par The use of machine learning to facilitate the error mitigation and localization started from the early 2000s.  In \cite{vigneau}, the authors proposed method for multipath effect mitigation using a hybrid neural network architecture based on multilayer perceptron for Law Earth Orbit (LEO) satellites. Otherwise, \cite{phan} treats the case of multipath mitigation using a Support Vector Regression (SVR) model. Here, the author integrates kernel support vector regressor with geometrical features to deal with the code multipath error prediction on ground fixed GPS stations. Subsequent works were dedicated to the selection of different types of the relevant features to detect and mitigate multipath in their application. Examples of such work can be seen in \cite{suzuki} and \cite{hsu} for non-line of sight (NLOS) multipath detection where the features are directly extracted from correlator output. In \cite{quan}, the authors give the detailed description of CNNs and their application in the problem of carrier-phase multipath detection. In this work, the feature map was extracted from multivariable time series from the end of signal processing stage using a 1D convolutional layers.
\par Deep learning models in GNSS systems were also used for the problem of spoofing attack detection in GNSS receiver \cite{spoof}. In this work, early-late phase, delta and signal level are the three main features extracted from the correlation output of the tracking loop. Using these features, spoofing detection is carried out using Deep Neural Networks (DNN) with fully-connected layers within short detection time.
\par CNN have been used also for GNSS integration with misalignement detection of Inertial Navigation System (INS) \cite{su}. The CNN was trained to detect the optical flow from smartphone camera to estimate the angle difference from moving direction to the inertial sensor. It allows to represent the distribution of objects moving on different images for misalignement calculation. 
\par Our proposed method also uses a CNN to automatically extract relevant feature maps from the GNSS correlator output. Input data are represented in the form of 3D tensors in time-frequency domain. We discuss in details this idea in the next sections. 
\section{Methodology}
\subsection{GNSS receiver architecture and signal processing}
\par
\begin{figure*}
	\centering
	\includegraphics[scale=0.6]{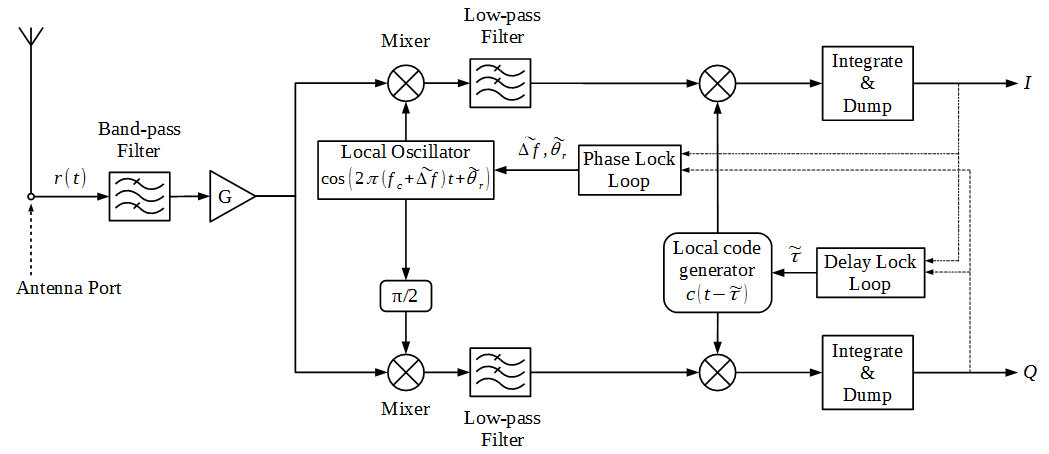}
	\caption{Typical model of the signal correlation chain of a GNSS receiver}
	\label{capteur_gnss}
\end{figure*}

 In this section we will briefly present the model of GNSS correlator output. In this work, we consider the simplified model of signal at the antenna port, as can be seen in Figure \ref{capteur_gnss}, which can be represented as:
\begin{equation*}
    r(t) = s(t) + b(t)
\end{equation*}
The desired signal $s(t)$ is modeled as follows \cite{kaplan2005understanding}:
\begin{equation*}
    s(t) = \sqrt{2C} D(t-\tau) c(t-\tau) \cos(2\pi (f_c + \Delta f)t + \theta_r)
\end{equation*}
where
\begin{itemize}
    \item $C$ is the power of received signal;
    \item $D(t)$ is the navigation message;
    \item $\tau$ is the propagation delay;
    \item $c(t)$ is the Pseudo-Random Noise (PRN) code sequence, corresponding to a specific satellite;
    \item $f_c$ is the carrier frequency;
    \item $\Delta f$ is the doppler offset
    \item $\theta_r$ is the carrier phase.
\end{itemize}
In this simplified model, we have taken the following hypothesis: 
\begin{itemize}
    \item The $D(t)$ and $c(t)$ terms are not distorted by propagation;
    \item The propagation delay $\tau$ and Doppler offset $\Delta f$ are constant;
    \item The propagation delay $\tau$ and carrier phase $\theta_r$ are supposed to be independent.
\end{itemize}
The model of noise is considered to be the Additive White Gaussian Noise (AWGN), limited in frequency by the bandwidth of RF front-end. Indeed, it is the model of the noise as seen at the input of the RF front-end stage transfered to the antenna output. The noise term $b(t)$ can be replaced by its Rice representation \cite{rice} as follows:
\begin{equation*}
    b(t) = n_i (t) cos(2\pi f_{lo}t + \theta_r) - n_q (t) sin(2\pi f_{lo}t + \theta_r)
\end{equation*}
where
\begin{itemize}
    \item $b(t) \sim \mathcal{N}(0, \NZERO B_{rf})$, AWGN, white at least on the band of RF front-end;
    \item $n_q (t) \sim \mathcal{N}(0, \NZERO B_{rf})$ and $n_q (t) \sim \mathcal{N}(0, \NZERO B_{rf})$ are AWGN on the bandwidth $B_{bb}=\frac{B_{rf}}{2}$;
    \item $\frac{\NZERO}{2}$ is the double sided noise Power Spectral Density (PSD);
    \item $B_{rf}$ is the RF front-end bandwidth. 
\end{itemize}
The aim of the receiver is to estimate the propagation delay $\tau$ for each satellite in view so as to calculate the user position according to a trilateration principle. $\Delta f$ and $\theta_r$ are nuisance parameters which have to be estimated too. 
\par The first step in the signal processing chain is the multiplication of the incoming signal by two local replicas in quadrature of the carrier signal. This operation splits the signal into two channels: in-phase (I) and quadrature (Q) as follows:
\begin{equation*}
    p_{\cos}(t) = r(t) \cdot \cos(2\pi (f_{c} + \tilde{\Delta f})t + \tilde{\theta_{r}})
\end{equation*}
\begin{equation*}
    p_{\sin}(t) = -r(t) \cdot \sin(2\pi (f_{c} + \tilde{\Delta f})t + \tilde{\theta_{r}})
\end{equation*}
where $\tilde{\Delta f}$ is the local estimate of the doppler offset $\Delta f$ and $\tilde{\theta_r}$ is the local estimate of the initial phase $\theta_r$ of the carrier signal. The two components $p_{\cos}(t)$ and $p_{\sin}(t)$ are then low-pass filtered to remove the high-frequency terms in $2f_{c}$.
\par Then, the two channels are correlated with a locally generated PRN code sequence. This correlation operation is performed by the means of a multiplier followed by the Integrate-and-Dump (I\&D) stage. The latter is parametrized by the coherent integration time $T_i$. 
\par Finally, a model of the signal available at the output of the I\&D stage, which is the correlator output, is \cite{handbook_gnss}:
\begin{eqnarray} \label{voie_i}
I = M \cos(\pi (\Delta f - \tilde{\Delta f})T_i + (\theta_r - \tilde{\theta_r}))  \notag \\ 
    \times \sinc(\pi (\Delta f - \tilde{\Delta f}) T_i) + n_I
\end{eqnarray}
\begin{eqnarray} \label{voie_q}
    Q = - M \sin(\pi (\Delta f - \tilde{\Delta f})T_i + (\theta_r - \tilde{\theta_r})) \notag \\ 
    \times \sinc(\pi (\Delta f - \tilde{\Delta f}) T_i) + n_Q
\end{eqnarray}
where 
\begin{itemize}
    \item $M = \frac{D T_i}{2} \sqrt{\frac{C}{2}} K(\tilde{\tau} - \tau)$
    \item $D$ is the value of the bit of the navigation message, which remains constant over the coherent integration time $T_i$;
    \item $K(\tilde{\tau} - \tau)$ is the auto-correlation function of the PRN code in $\tilde{\tau} - \tau$;
    \item $\tilde{\tau}$ is the local estimate of the propagation delay $\tau$, hence $\tilde{\tau} - \tau$ is the propagation delay estimation error;
    \item $\Delta f - \tilde{\Delta f}$ is the doppler estimation error;
    \item $\theta_r - \tilde{\theta_{r}}$ is the phase estimation error; 
    \item $n_I (t) \sim \mathcal{N}(0, \frac{\NZERO T_i}{16})$ and $n_Q (t) \sim \mathcal{N}(0, \frac{\NZERO T_i}{16})$ are two independent and identically distributed white noise components; 
\end{itemize}
It is worth noting that the $I$ and $Q$ components of the correlator output are being produced each $T_i$.
\subsection{CNN general description}
\par In this part, we will describe the motivation behind the application of CNN in the GNSS correlation output data processing. We try to take advantage of the fact that in the case of structured and hierarchical data CNNs are capable to efficiently extract feature maps. The correlator output signal can be seen as a structured and hierarchical data.
\par Indeed, the correlator output double-channel data can be easily represented in the form of a 3D tensor like multi-channel "image" and  correlator output values can be translated to the predictive model as the intensity of pixels in both I and Q channels. 
\par The definition and principles of CNN are described in details in \cite{chollet, goodfellow}. According to \cite{goodfellow}, convolution leverages three important ideas that can help improve a machine learning system: sparse interactions, parameter sharing and equivariant representations. Unlike the classical DNN where every output unit interacts with every input unit, through matrix multiplication of parameters, CNN typically promotes sparse interactions. This also means that fewer parameters need to be stored, which reduces the memory requirements of the model and improves its statistical efficiency and decreases its operation complexity.
\par The data used with CNNs usually consists of several channels, each channel being the observation of a different quantity in different axis scales. One of such data type is a color image data where each channel contains the color of pixels. The convolution kernel moves over both the horizontal and the vertical axes of the image, conferring translation equivariance in both directions. 
\par
\begin{figure*}
    %\label{cnn_example_img}
    \begin{center}
    \includegraphics[scale=0.3]{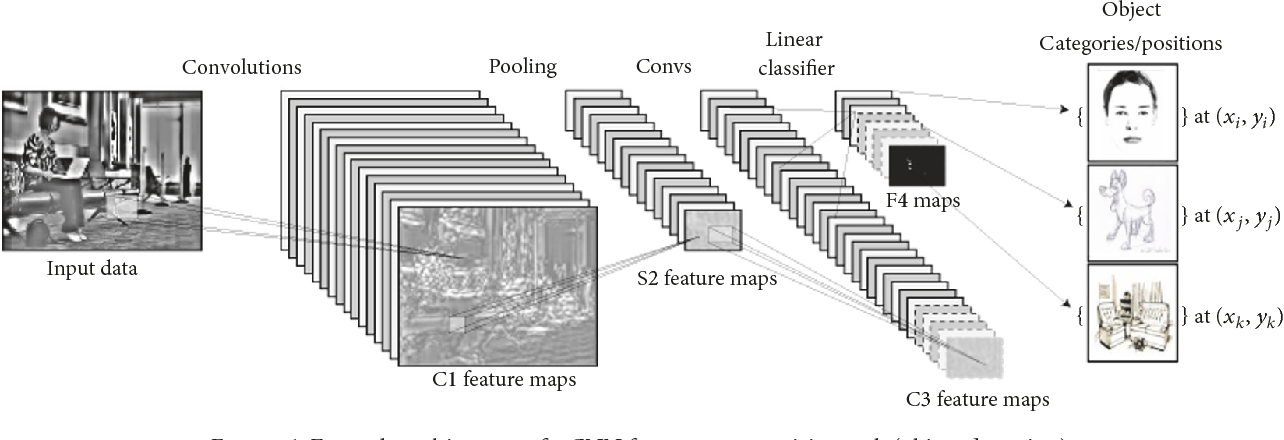}
    \caption{CNN structure and main elements example \cite{cnn_example}}
    \label{cnn_example_img}
    \end{center}
\end{figure*}

 A typical CNN architecture is composed of the following basic elements as it can be seen in Figure~\ref{cnn_example_img}.
\begin{itemize}
    \item \textbf{Convolutional layer:} A convolutional filter (kernel) within convolutional layers provides a compressed representation of input data. With convolutional filters computed using CNN, convolutional layers can extract features from input data. Each filter is composed of weights that are tuned during the training phase of the network.
    \item \textbf{Pooling layer:} The convolved features are subsampled by a specific factor in the subsampling layer. The role of a subsampling layer is to reduce the variance of convolved data so that the value of a particular feature over a region of an input layer can be computed and merged together. 
    \item \textbf{Activation function:} Following several convolutional and pooling layers, the high-level reasoning in the neural network is performed via fully connected layers. Neurons in a fully connected layer have full connections to all activation in the previous layer. Fully connected layers eventually convert the 2D feature maps into a 1D feature vector. The derived vector either could be fed forward into a certain number of categories for classification or could be considered as a feature for further processing.
\end{itemize}
The specific architecture and setup of our CNN model dedicated to the multipath detection at the correlator output will be described in the Section \ref{experiment_section}. 
\section{GNSS Data generation and labelling}
\subsection{Main signal generation}
\par
\begin{center}
\begin{figure}
    \includegraphics[scale=0.3]{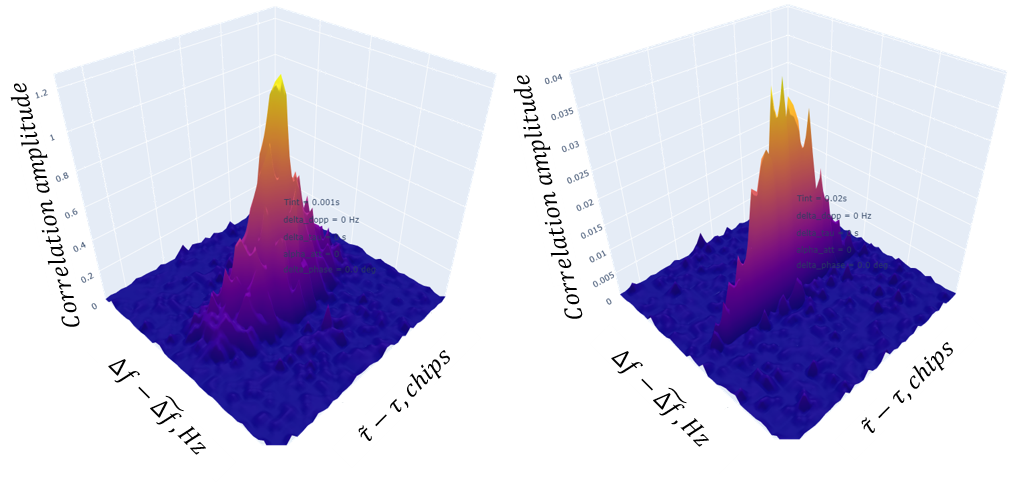}
    \caption{Synthetic correlator output visualization for $T_{int} = 1 \ms$ (left) and $T_{int} = 20 \ms$ (left)}
    \label{int_time_peaks}
\end{figure}
\end{center}

 In order to test our prediction models, an artificial signal generator was developed. The data is generated in the form of two matrices, one for each of the $I$ and $Q$ channels, according to equations (\ref{voie_i}) and (\ref{voie_q}). The axes of these matrices are in doppler estimation error $\Delta f -  \tilde{\Delta f}$ and code delay estimation error $\tau - \tilde{\tau}$. The output data corresponding to this main signal can be parametrized as follows:
\begin{itemize}
    \item Coherent integration time $T_i$ in $\ms$;
    %\item Doppler and code delay offsets for the reflected signal due to the multipath effect.
    \item Carrier-to-noise ratio \CNZERO in $\dBHz$. 
\end{itemize}
An illustration of the output of our generator in terms of $I(t)^2 + Q(t)^2$ is given in Figure \ref{int_time_peaks} for two different integration times.
\begin{figure*}
    \begin{center}
    \includegraphics[scale=0.35]{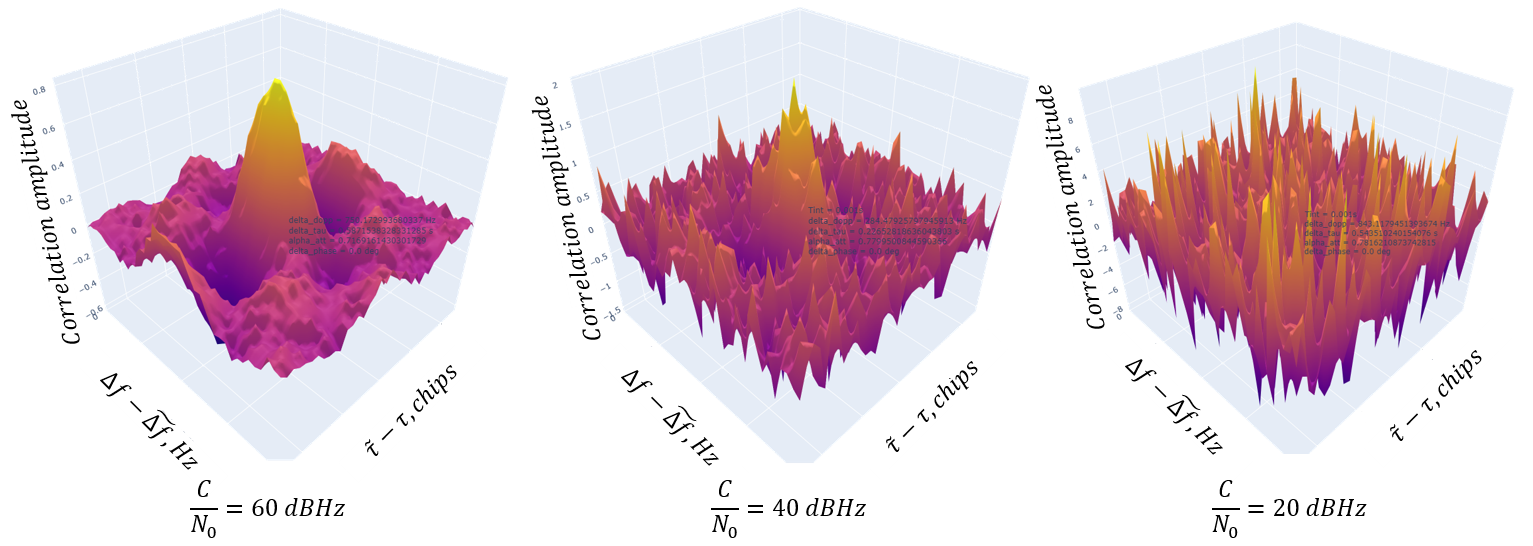}
    \caption{Synthetic correlator output visualization for I-channel for different values of carrier-to-noise ratio}
    \label{cn0_peaks}
    \end{center}
\end{figure*}
\subsection{Multipath generation}
\par If a multipath signal is received in addition to the main signal, as the signal processing chain is linear, we can then consider that the correlator output is the sum of the correlator output of the main signal plus the one due to the multipath: 
\begin{eqnarray*}
    I' &=& I + I_{MP} (\alpha_{MP}, \Delta \tau_{MP}, \Delta f_{MP}, \Delta \theta_{MP}) \\
    Q' &=& Q + Q_{MP} (\alpha_{MP}, \Delta \tau_{MP}, \Delta f_{MP}, \Delta \theta_{MP})
\end{eqnarray*}
Where
\begin{itemize}
    \item $\Delta \tau_{MP} > 0$ is the code delay in excess to the main signal delay $\tau$;
    \item $\Delta f_{MP}$ is the difference between the doppler offset of the main signal and the multipath;
    \item $\Delta \theta_{MP}$ is the difference between the phase of the main signal and the multipath;
    \item $\alpha_{MP}$ is the mulipath attenuation coefficient in comparison to the main path. 
\end{itemize}
The generation range for the code delay estimation error is set to $[0,2]\chips$, where one $\chip$ is the duration of one bit of the PRN code sequence. This is because outside of this interval the correlation value is negligible, so the multipath effect is annihilated. Regarding the doppler offset estimation error range, it is defined as $\pm\frac{1}{T_i}$, the width of the main lobe of the $\sinc$ function in (\ref{voie_i}) and (\ref{voie_q}). We have considered that outside of this interval the side lobes of the $\sinc$ function attenuate enough the multipath component. 
\subsection{Training Dataset construction}
\par Also it needs to be noticed that to be represented in the form of 3D tensors, the processed signal needs to be discretized by a predefined number of correlator outputs depending on the given GNSS receiver configuration. So to cover all such possible configurations we have used the following set of discretization levels: $N \in \{4, 6, 8, 10, 20, 30, 40\} $ points in each range $\tau - \tilde{\tau}$ and $\Delta f - \tilde{\Delta f}$. That is to say the number of correlator outputs is $N^2$.
\par As the multipath effect cannot be considered as a rare event (for example in urban environment), the simulated dataset was generated so that it is well balanced between two classes:\\
\begin{itemize}
    \item Category A: represents normal signal with no reflections (data from Category A is labelled as direct-signal only data).
    \item Category B: simulates signals with a single multipath effect which can include doppler, code delay or phase offsets (data from Category B is labelled as multipath event).
\end{itemize}
See \cite{github} for a public release of our generator code.
\section{Experiment} \label{experiment_section}
\subsection{Selected CNN architecture}
\par The architecture and training parameters of the designed CNN are shown in Table \ref{hyperparam_cnn}.
\par The feature extraction part of the CNN consists of 4 convolutional blocks with 2 convolutional layers in each block and pooling layers at the end of each block. For example, consider a correlator output image cropped up to discretization size $N = 40 \times 40 \times 2$ that is passed through the network. The first convolutional layer from the first block (conv1-1) is a feature map with a size $40 \times 40 \times 16$, where the dimensions indicates height, width and number of channels. This layer is generated by the convolutional operation of 16 filters. Each filter corresponds to a channel of the feature map, respectively. By using maxpooling operation of pooling size $2 \times 2$ and stride size $2 \times 2$, the dimensions feature of the map are reduced to $20 \times 20 \times 16$ in the second convolutional layer of the first block (conv1-2).
\par The decision to use the convolutional layers assembled in blocks was made to avoid the use of large-sized kernels (for example $7 \times 7$ or $5 \times 5$) with multiple $3 \times 3$ kernel-sized filters one after another. It is especially true for the case of scalable image discretization size which can vary in wide range from 40 to 300 pixels. This kind of architecture was succesfully applied in the VGG family of convolutional neural networks \cite{vgg} on the ImageNet dataset of over 14 million images belonging to 1000 classes.Thus VGGNet-like architectures can be considered as a baseline feature extractors.
\par The classifier part of the CNN consists of two fully-connected layers of size $1152 \times 1$ and $256 \times 1$ with the sigmoid output activation function.
\begin {table}
    \begin{center}
    \caption {CNN hyperparameters}
        \begin{tabular}{ rrr } 
            \hline
            Nb Layers & 4 Blocks of 2 Conv layers \\
            Loss & LogLoss \\ 
            Optimizer & Adam \\
            Learning rate & $10^{-3}...10^{-4}$ \\
            Batch size & 32 \\
            Nb epochs & 30 \\ 
            \hline
        \end{tabular}
    \label{hyperparam_cnn}
    \end{center}
\end {table}
\subsection{Benchmark model}
\par The performance of the proposed method is compared to those described in \cite{suzuki}. The method is based on the Support Vector Mashine (SVM) approach. Reported results show that 87\% of the signals were correctly discriminated (results depend on the number of correlation data points used). Since the method also collects data from the output of the correlator block, it follows the same process as our technique.
\par We have implemented the same feature extraction pipeline and used the same SVM classifier hyperparameters \cite{Vapnik1998, Scholkopf:2001:LKS:559923} on our synthetic dataset to compare the efficiency of the proposed CNN algorithm (referred as MultipathCNN in the following). To obtain the correlation shape we also use 13 correlator outputs.
\par As in \cite{suzuki}, the following features were extracted: 
\begin{itemize}
    \item Number of local maxima of the correlation outputs per period $F_2 =  \frac{N_{local-maxima}}{\Delta t}$;
    where $\Delta t$ is the correlation interval taken equal to coherent integration period.
    \item Distribution of the delay of the maximum correlation output $F_3 = \frac{1}{M} \sum_{i=1}^{M} (t_{i-max} - \bar{t})^2$ where $t_{i-max}$ is the code delay of the maximum correlation output, $\bar{t}$ is the mean of the code delay, and $M$ is the number of correlator output samples.
\end{itemize}
In \cite{suzuki}, the authors have also used a  signal strength vs. elevation angle feature (referred as $F_1$) that is not taken into account here as we have not introduced the physical context of experiments (receiver's speed, satellite constellation) since the generated dataset is synthetic. Thus, we have made the choice to not use the feature $F_1$ as opposed to \cite{suzuki}.
\par Table \ref{hyperparam_svm} reports the SVM hyperparameters that were used during the experiment.
\begin {table}
    %\begin{tabular}
    \begin{center}
    \caption {SVM hyperparameters}
        \begin{tabular}{ rrr } 
            \hline
            Kernel & 4 RBF (Gaussian) \\
            Cross Val & 3 Folds \\ 
            Normalization & Standard Scale \\
            \hline
        \end{tabular}
    \label{hyperparam_svm}
    \end{center}
    %\end{tabular}
\end {table}
\subsection{Experimental setup}
\par The strategy of experiments was designed for the cases of coherent integration time of $T_i \in \{1,20\} \ms$. The lower value corresponds to the receiver operation prior to synchronisation with the navigation message, the largest one being used after synchronisation. For each of two values of integration time we have conducted two types of tests: 
\begin{itemize}
    \item Tests on various discretization levels $N$;
    \item Tests on models benchmark comparison.
\end{itemize}
It is worth noting that for both types of test above, we make comparisons for several values of the following parameters: 
\begin{itemize}
    \item $\Delta \theta_{MP} = \{0, 45, 90, 180\}^\circ$;
    \item $\CNZERO \in \{20, 30, 40, 50, 60\} \dBHz$;
    \item $\alpha_{MP} \sim \mathcal{U}([0.5, 0.9])$.
\end{itemize}
A variation of the $I$ component with respect to \CNZERO is illustrated in Figure \ref{cn0_peaks}.
\subsubsection{Type 1. Tests on various discretization levels}
During this stage we have conducted tests of our MultipathCNN algorithm on discretization levels $N \in \{4, 6, 8, 10, 20, 30, 40\}$ to estimate the optimal value of discretization. To model the multipath effect we have picked randomly the doppler offset uniformly as $\Delta f_{MP} \sim \mathcal{U}([0,1000])\Hz$  and the code delay offset as $\Delta \tau_{MP} \sim \mathcal{U}([0.1...0.8]) \chips$ separately for each discretization level. 
\subsubsection{Type 2. Test on models benchmarks comparison}
In this stage, we have performed the test to compare the performance of our MultipathCNN model with the reference SVM model. We have chosen fixed discretization level of 10 correlator outputs to be closer to the case of SVM (with its 13 correlator outputs). We have also picked randomly and uniformly the doppler and the code delay offsets in the same range as in the first type of experiments with simultaneous variation of these parameter.
\begin{figure*}
	\centering
	\includegraphics[scale=0.5]{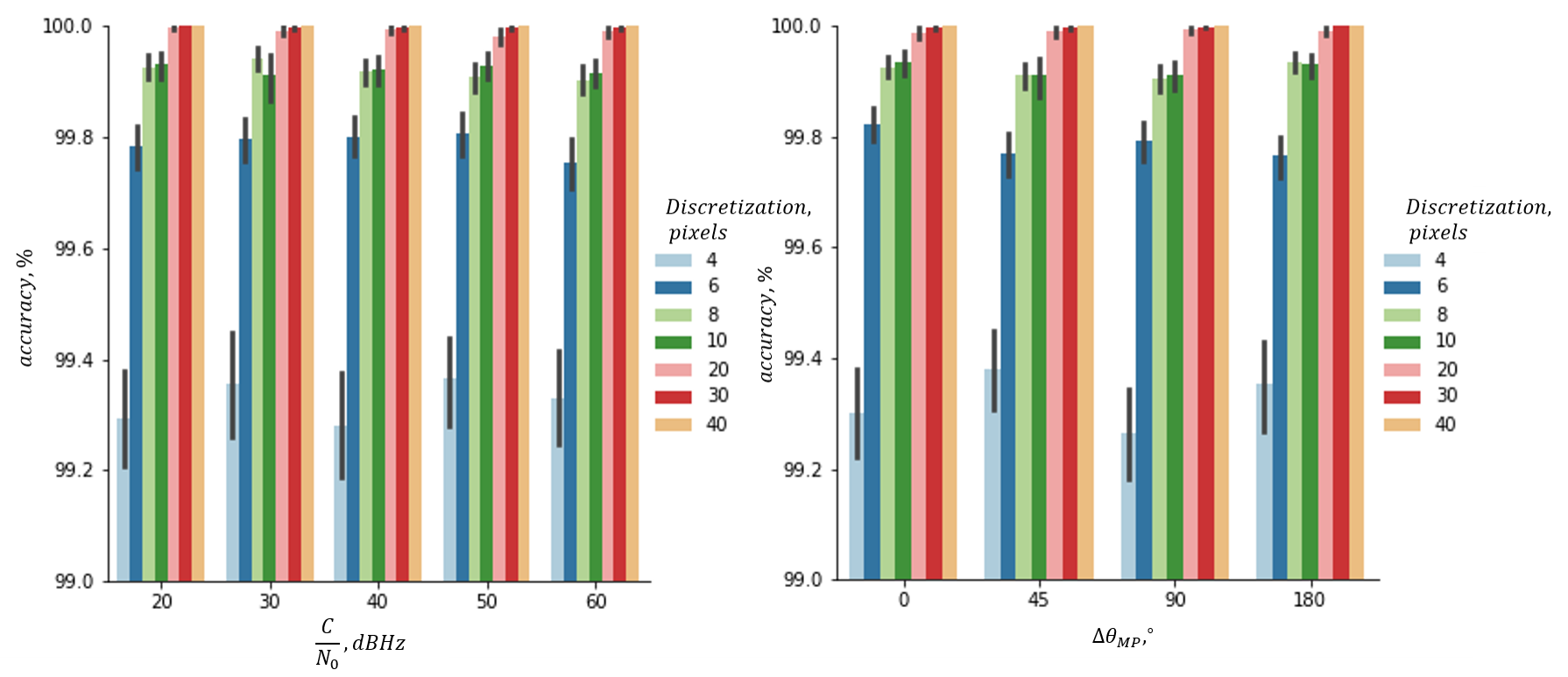}
	\includegraphics[scale=0.5]{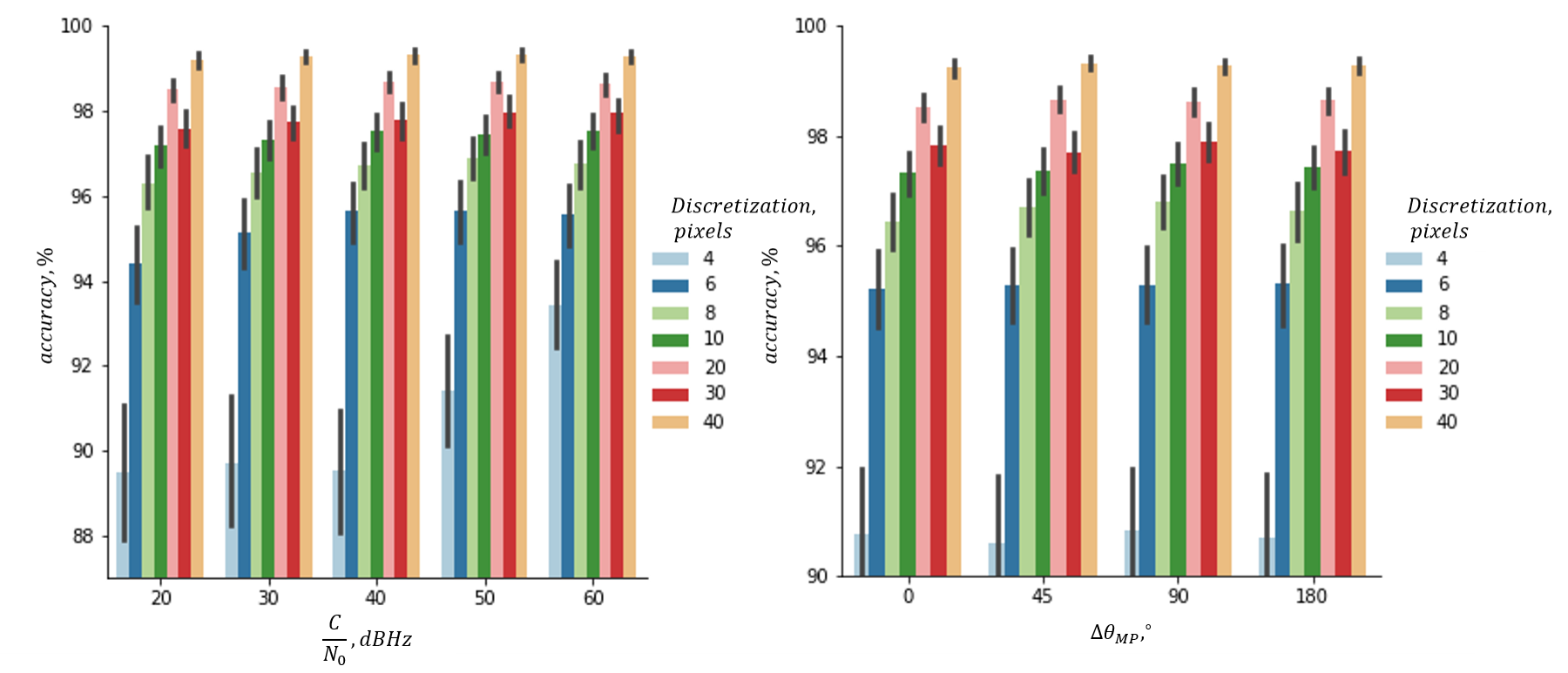}
	\caption{Accuracy in \% averaged over 20 runs of MultipathCNN for various levels of discretization for two integration times: 1 $\ms$ (up) and 20$\ms$ (bottom)}
	\label{discr_test}
\end{figure*}
\subsection{Results tables and figures description}
\par In order to evaluate the performance of multipath detectors, the two models were applied directly to the synthetic dataset which was varying in different configurations listed above. We compare our results with state-of-the-art method using an accuracy metric.
\par In Figure~\ref{discr_test} we provide the barplots which represent the accuracies of the model MultipathCNN on prediction of multipath. These results are averaged over 20 runs of the test of type 1 described earlier. Here the results are represented as function of different $\CNZERO$ values and multipath phase offsets $\Delta \theta_{MP}$. For each of these parameters the results are also grouped by the discretization level $N$.
\par In tables \ref{model_test_phase} and \ref{model_test_cn0}  we provide also the values of accuracy averaged over 20 runs for different levels of $\CNZERO$ and $\Delta \theta_{MP}$. These results correspond to the tests of type 2 described above. \\
\begin {table}
    \begin{center}
    \caption {Average model performance on benchmark test as the function of phase offset}
        \begin{tabular}{ rrr } 
            \hline
            \multicolumn{3}{c}{$T_i = 1 \ms$}\\
            \hline
            $\Delta \theta_{MP}$ & SVM, \% & MultipathCNN, \% \\
            \hline
            0, $^{\circ}$ & $69.9\pm20.8$ & $\textbf{99}\pm0.1$\\
            45, $^{\circ}$ & $69.5\pm20.5$ & $\textbf{99.9}\pm0.1$\\
            90, $^{\circ}$ & $69.5\pm21$ & $\textbf{99.8}\pm0.2$\\
            180, $^{\circ}$ & $69.9\pm21.3$ & $\textbf{99.9}\pm0.2$\\
            \hline
            \hline
            \multicolumn{3}{c}{$T_i = 20 \ms$} \\
            \hline
            $\Delta \theta_{MP}$ & SVM, \% & MultipathCNN, \% \\
            \hline
            0, $^{\circ}$ & $93.6\pm13.2$ & $\textbf{94.7}\pm1.2$\\
            45, $^{\circ}$ & $93.5\pm13.5$ & $\textbf{94.7}\pm1.2$\\
            90, $^{\circ}$ & $93.1\pm14$ & $\textbf{94.6}\pm1.3$\\
            180, $^{\circ}$ & $93.5\pm13.2$ & $\textbf{94.7}\pm1.2$\\
            \hline
        \end{tabular}
    \label{model_test_phase}
    \end{center}
\end {table}
\begin {table}
    \begin{center}
    \caption {Average model performance on benchmark test as the function of \CNZERO}
        \begin{tabular}{ rrr } 
            \hline
            \multicolumn{3}{c}{$T_i = 1\ms$}\\
            \hline
            $\CNZERO, \dBHz$ & SVM, \% & MultipathCNN, \% \\
            \hline
            20 & $62.2\pm14.1$ & $\textbf{99.8}\pm0.2$\\
            30 & $73.8\pm23.9$ & $\textbf{99.9}\pm0.1$\\
            40 & $71.1\pm21.3$ & $\textbf{99.9}\pm0.2$\\
            50 & $70.7\pm21$ & $\textbf{99.8}\pm0.2$\\
            60 & $70.8\pm21.1$ & $\textbf{99.8}\pm0.2$\\
            \hline
            \hline
            \multicolumn{3}{c}{$T_i = 20\ms$} \\
            \hline
            $\CNZERO, \dBHz$ & SVM, \% & MultipathCNN, \% \\
            \hline
            20 & $67.5\pm7.7$ & $\textbf{94.3}\pm1.2$\\
            30 & $\textbf{99.6}\pm1$ & $94.4\pm1.1$\\
            40 & $\textbf{100}\pm1.1$ & $94.8\pm1.2$\\
            50 & $\textbf{100}\pm1.1$ & $94.8\pm1.1$\\
            60 & $\textbf{100}\pm1$ & $95\pm1.2$\\
            \hline
        \end{tabular}
    \label{model_test_cn0}
    \end{center}
\end {table}
\subsection{Discussion on results}
\subsubsection{Tests on various discretization level}
\par As it can be seen in Figure~\ref{discr_test}, the reduction of discretization level up to 4, 6 discretization levels leads in average to the reduction of model accuracy by 8\% and degrades slightly the variance of other MultipathCNN model accuracy. It can also be seen that for the case of integration time of 20 $\ms$, the models with 4 and 6 correlator outputs become sensitive to the $\CNZERO$ variation. However, for the case of integration time of 1 $\ms$, $\CNZERO$ variation has almost no significant influence on the models performance. Regarding the influence of the phase offset, the sole case of $\Delta \theta = 90^{\circ}$ shows the degradation in the model metrics by at most 0.5\%.
\par As our final objective in the fututre is to integrate the multipath detector in the GNSS receiver, we need to find the minimum number of correlator outputs $N$ keeping an accuracy high enough to detect the multipath effect. Taking into account the test results, $N \in \{8, 10\}$ seems a reasonable value which is insensitive to the level of noise in the signal.
\subsubsection{Tests on models benchmark comparison}
\par
\begin{figure*}
	\centering
	\includegraphics[scale=0.5]{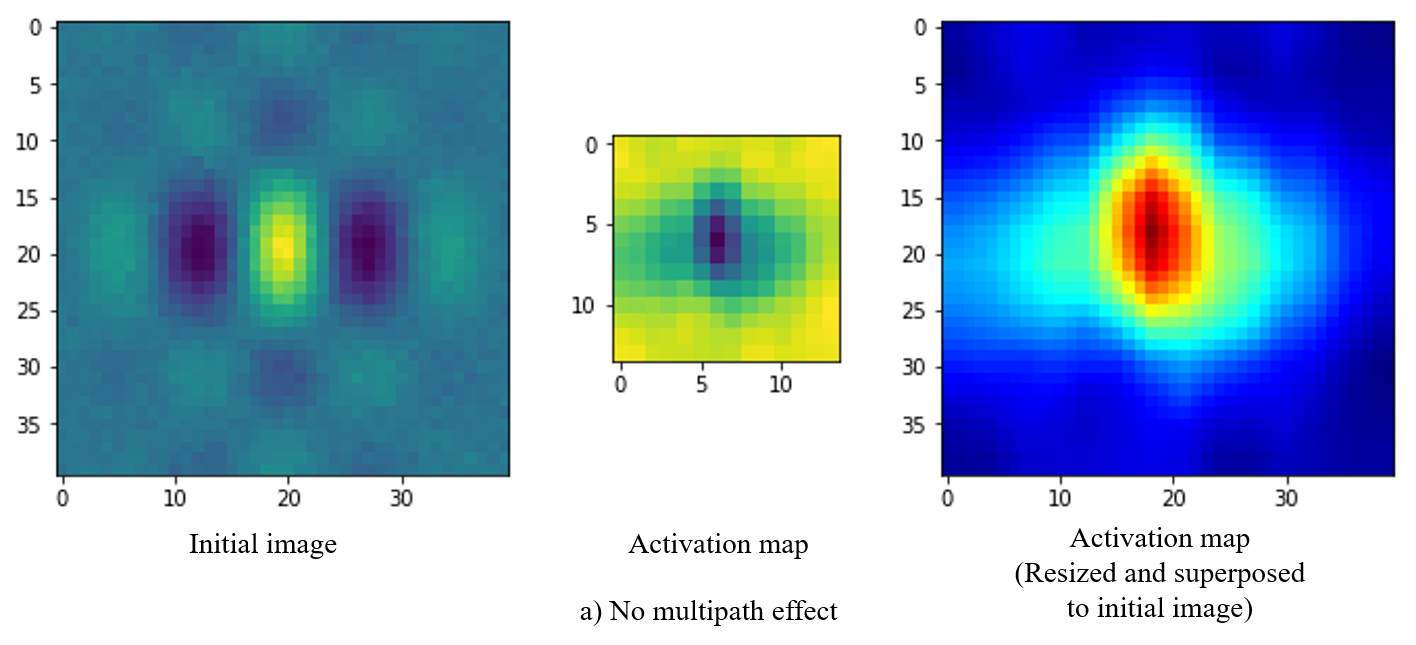}
	\caption{Activation map from the last convolutional layer of the MultipathCNN for the case without multipath in the tensor with discretization $40 \times 40$ for integration time $T_i=1\ms$}
	\label{heatmap_nomp}
\end{figure*}

\begin{figure*}
	\centering
	\includegraphics[scale=0.5]{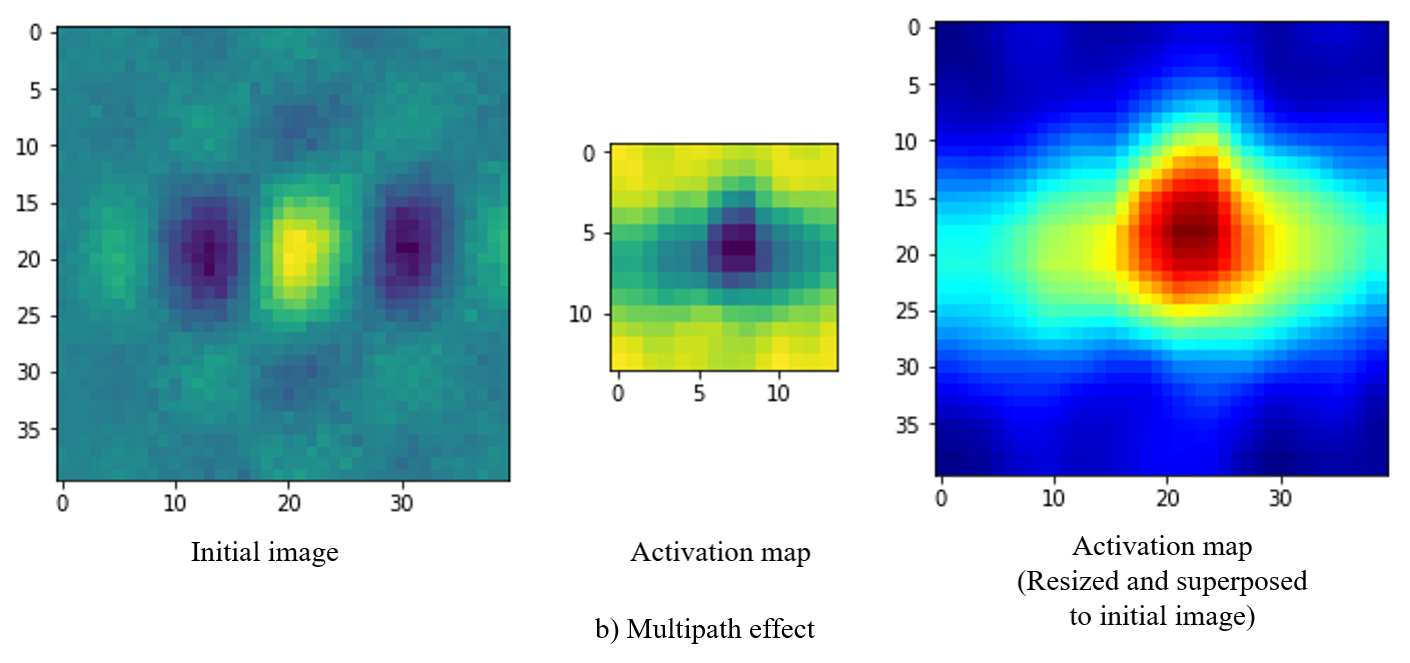}
	\caption{Activation map from the last convolutional layer of the MultipathCNN for the case with multipath in the tensor with discretization $40 \times 40$ for integration time $T_i=1\ms$}
	\label{heatmap_mp}
\end{figure*}

 The results shows that our proposed method outperforms the SVM on the synthetic GNSS dataset. Depending on the coherent integration time, the results are represented as follows:
\begin{itemize}
    \item For the case of integration time of $T_i = 1 \ms$, our model outperforms the SVM on every value of \CNZERO and phase offset. 
    \item For the case of integration time of $T_i = 20 \ms$ and the values of $\CNZERO \in [30, 60] \dBHz$ our model shows 5\% less accuracy than the benchmark, but outperforms it for low $\CNZERO$ value.
\end{itemize}
We believe that these performance results of the MultipathCNN are due to the fact that the CNN is able to catch the geometrical dependencies in the data. To demonstrate this property, we have used the visualized heatmaps of class activation at the last convolutional layer. This can help to understand which part of an image led a CNN to its final classification decision. As it can be seen on Figures \ref{heatmap_nomp} and \ref{heatmap_mp}, the activation map of the last convolutional layer of MultipathCNN shows that the region of importance of feature map of the model is slightly distorted in the case of multipath. 
\section{Conclusion and further work}
\par In this study a CNN based multipath detection method has been proposed to detect multipath in the GPS L1 C/A correlator output stage. The proposed method has been validated with synthetic GPS data generator. The performance of the detector has been compared to SVM based multipath detector. The results have shown that MultipathCNN performs better than the compared method especially when the \CNZERO degrades.
\par The results also show that the proposed MultipathCNN technique is well suited to the multipath detection task. Further experiments with physical signals should be  conducted to confirm its applicability in real life multipath situation.
\section*{Acknowledgements}
\par This project has been partly funded by the SESAR Joint Undertaking under the European Union's Horizon 2020 research and innovation programme under grant agreement No 783287. The opinions expressed herein reflect the authors' view only. Under no circumstances shall the SESAR Joint Undertaking be responsible for any use that may be made of the information contained herein. 
\bibliographystyle{IEEEtran}
\bibliography{biblio_ref_10_14}

\end{document}